\documentclass[twocolumn,prb]{revtex4-2}
\usepackage{amsmath,amssymb,graphicx,color,hyperref}
\usepackage{bm}
\usepackage{amsfonts}
\usepackage[T1]{fontenc}
\usepackage{amsmath,amsbsy,amssymb,graphicx}
\usepackage{times}
\usepackage{subfigure}
\usepackage{amsthm}
\usepackage{lipsum}

\def\RR{\mathbb{R}}

\newcommand{\be}{\begin{equation}}
	\newcommand{\ee}{\end{equation}}
\newcommand{\bea}{\begin{eqnarray}}
	\newcommand{\eea}{\end{eqnarray}}

\newcommand{\ed}{\end{document}}

\newcommand{\bi}{\begin{itemize}}
\newcommand{\ei}{\end{itemize}}

\newcommand{\bce}{\begin{center}}
\newcommand{\ece}{\end{center}}

\begin{document}

\title{A class of stable nonlinear non-Hermitian skin modes}

\author{Hamed Ghaemi-Dizicheh}
\email {hamed.ghaemidizicheh@utrgv.edu}
\affiliation{$^1$Department of Physics and Astronomy, University of Texas Rio Grande Valley, Edinburg, Texas 78539, USA
}


\begin{abstract}
The non-Hermitian skin effect (NHSE) is a well-known phenomenon in open topological systems that causes a large number of eigenstates to become localized at the boundary. Although many aspects of its theory have been investigated in linear systems, this phenomenon remains novel in nonlinear models. In the first step of this paper, we look at the conditions for the presence of quasi-skin modes in a semi-infinite, one-dimensional, nonlinear, nonreciprocal lattice. In the following phase, we explore the survival time of the quasi-skin mode in a finite nonlinear lattice with open edges. We study the dependency of the survival time on the system's parameters and demonstrate how the nonreciprocity of the system affects the survival time. This study introduces a method for achieving a stable localized state in a nonlinear finite lattice.

\end{abstract}
\maketitle
\section{introduction}
In recent years, there has been a lot of interest in a novel form of topological tight-binding systems with active gain and loss or nonreciprocal couplings \cite{schomerus2013topologically,Rud09,Zhu14,Lee16,ashida2020non,Zeng2020PRB,Liu2020PRB,Kunst2018skin,Zhou2020PRB,ghaemi2021compatibility,2023 general}. The so-called non-Hermitian systems introduced unique physics and phenomena with applications in a wide range of physical realms. One of the prominent phenomena in the non-Hermitian topological models absent in the Hermitian one is the skin effect \cite{Yao2018skin,Lee2019skin,Lee2019PRLskin,Borgina2020skin,lee2016anomalous,yao2018edge,lieu2018topological} where eigenstates of a lattice with open boundaries exhibit localized behaviors leading to failure of the bulk-boundary correspondence due to such a nonlocal change of the eigenstates. The main characteristic in non-Hermitian models that leads to a skin state is non-reciprocity, where the amplitudes of the right-going and left-going couplings differ. Recently, another intriguing non-Hermitian phenomenon has been introduced such that a substantial portion of loss occurs at the system boundary known as non-Hermitian edge burst \cite{Xue2022edgeburst}.\\
Moreover, nonlinearity could play an important role in different topological platforms, including mechanical \cite{SNEE2019100487,Lo2021mechanical}, photonic \cite{Leykam2016photonic,Maczewsky2020photonic,kruk2019nonlinear,kirsch2021nonlinear,righini2023advances}, electric circuit \cite{Ezawa2022electriccircuit}, and resonator \cite{Zangeneh2019resonator} models. Hence, one can investigate non-Hermitian extensions of such models to discover novel phenomena such as nonlinear skin effect.\\ A nonlinear extension of the skin effect has been studied both from stationary and dynamic viewpoints. In Ref. \cite{Yuce2021PLA}, a stationary skin mode of a nonlinear non-Hermitian system with unidirectional coupling is studied under open boundary conditions (OBCs) and semi-infinite boundary conditions (SIBCs). The main characteristic that rises from nonlinearity is the emergence of a fractal spectrum in addition to the continuum one, where the localized mode happens in the continuous spectrum. The dynamics of the non-Hermitian skin effect and topological trap-skin phase have been explored in \cite{Ezawa2022PRB,Ezawa2022PRResearch} by making use of the quench method. The trap-skin state is formed in the strong nonlinearity regime where the pulse is trapped at the initial site coincident with the skin state. The condition on the existence of topological edge solitons is examined in a nonlinear Su-Schrieffer-Heeger (SSH) model with gain and loss using an analytic technique \cite{Bocharov2023Chaotic}. Recently, in a weak nonlinear regime, the dynamics of non-Hermitian skin mode has been investigated in nonlinear Hatano-Nelson model \cite{manda2023skin}.  \\
In this paper, we look at quasi-skin modes and investigate the time intervals across which these modes can maintain their initial profile in a truncated lattice. This time interval is referred to as survival time. The quasi-skin modes are the stationary solutions of the nonlinear Schr\"{o}dinger equation under semi-infinite boundary conditions, assuming a lattice with a boundary on the left but no boundary on the right. Indeed quasi-skin modes are a class of solutions among an extensive number of nonlinear solutions localized at the left edge and decrease monotonically by increasing the site number of the lattice. These solutions, in general, do not meet OBCs. However, for a large enough system, one can apply these solutions to approximately satisfy OBCs up to a survival time.\\ In other words, if a finite lattice is initially prepared with the skin mode of a semi-infinite lattice, that state will retain its spatial form until a survival time has passed. The linear version of the quasi-skin modes has been studied recently in  \cite{Cem2022stability}, where the survival time can be manipulated by changing the coupling at the end of the lattice. \\
Here, in the first step, we explore a condition on the system's parameters leading to the localized mode in a nonlinear nonreciprocal semi-infinite lattice. By satisfying this condition, one can obtain class of quasi-skin solutions. We then investigate the dynamics of these modes in a finite truncated system and find their survival time. We demonstrate how the system's parameters influence the survival time of the quasi-skin mode. For example, the strenght of nonreciprocity can magnify the influence of nonlinearity in making the quasi-skin mode unstable. On the other side, enlarging the system can result in a longer survival time for the quasi-skin mode. This study can open a pathway for studying stable non-Hermitian skin modes, which has the potential for application in topological laser \cite{Zhu2022PRL}.
\section{Model}\label{sec1}
We consider a one-dimensional non-Hermitian tight-binding model described by the following discrete nonlinear Schr\"{o}dinger equation
\be
i\dfrac{d\Psi_n}{dt}+\kappa_R\Psi_{n+1}+\kappa_L\Psi_{n-1}+g\vert\Psi_n\vert^{2}\Psi_n=0,\label{dnamical SE}
\ee
where $ n=1,2,\cdots,N $ indexes the lattice sites, $\Psi_n$ is the complex-valued field amplitude at the site $n$ as shown in Fig. \ref{SS for Semi-Infinite} (inset). Here, $ \kappa_R $ and $ \kappa_L $ are the forward and backward coupling amplitudes. One can parametrize the coupling amplitudes such that
\begin{align}
	&\kappa_R=\kappa(1+\lambda),&&\kappa_L=\kappa(1-\lambda),
\end{align}
where $ 0\leq\lambda\leq1 $ is the nonreciprocity parameter. The forward and backward couplings are nonreciprocal for $\lambda\neq0$, suggesting an asymmetric transport pattern. This is referred to as the non-Hermitian skin effect (NHSE) in the linear models, where all eigenstates are collected toward the left boundary under the OBCs. In an extreme case, $\lambda=1$, the forward coupling is turned off, and the coupling becomes unidirectional.\\ In this study, we assume a general Kerr-type (self-interaction) nonlinearity with strength $ g>0 $ applicable to numerous physical systems. The methodology and approach we used for positive-valued nonlinear strength can be directly extended to  negative-valued nonlinear strength. We focused primarily on the positive case for clarity and brevity but acknowledged that our approach remains valid for negative values.\\ The existence of nonlinear skin modes for the unidirectional coupling was explored in both open and semi-infinite boundary conditions \cite{Yuce2021PLA}. Here, we provide a condition for the existence of nonlinear quasi-skin modes for $\lambda\neq 1$ and investigate their time evolutions in an open system.
\subsection{A nonlinear quasi-skin mode}
\begin{figure}
	\begin{center}
		\includegraphics[scale=.6]{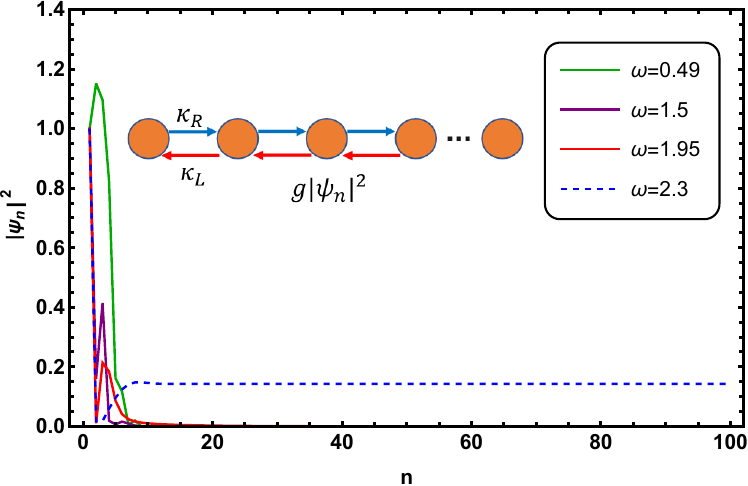}\hspace{0.5cm}
		\caption{The stationary solution of (\ref{stationary SE}) for SIBCs. In these diagrams, we take $ \kappa=1 $, $ \lambda=0.5 $, $ \vert\psi_{1}\vert=1 $, and $ g=2.1 $. Given these parameters, the nonlinear Schr\"{o}dinger equation admits skin mode for ($ 0.1<\omega<2 $). We show that the skin solution corresponds to the frequencies near the lower bound (blue line) and upper bounds (red line). Beyond this range, there is no localized state that decreases monotonically to zero (dashed blue line).}
		\label{SS for Semi-Infinite}
	\end{center}
\end{figure}

We are looking for the stationary solutions where the amplitude of the complex-valued field is defined by $ \Psi_n=e^{i\omega t}\psi_n $ with $ \omega $ being frequencies. There exist eigenstates with complex-valued eigenvalues, making the nonlinear term time-dependant, here in this letter, we restrict ourselves to the real-valued frequency $\omega\in\RR$. Then, we lead to the following set of nonlinear equations
\be
\omega\psi_n-\kappa_{R}\psi_{n+1}-\kappa_{L}\psi_{n-1}-g\vert\psi_{n}\vert^{2}\psi_{n}=0.\label{stationary SE}
\ee
To solve this problem, one can use either the SIBCs ($\psi_0=\psi_{\infty}=0$) or the OBCs ($\psi_0=\psi_{N+1}=0$). In the linear regime, skin modes can be derived analytically under OBCs, but in the nonlinear case, this is not practicable since the total number of solutions grows exponentially with the lattice size \cite{Yuce2021PLA}. In this paper, we take another approach to investigate nonlinear skin modes under OBCs. In this approach, we look at the class of skin modes under SIBCs known as the quasi-skin mode. It should be emphasized that each OBCs mode is a SIBCs mode, although the reverse is not necessarily true unless the system is large enough. However, we show that the solutions of Eq. (\ref{stationary SE}) under SIBCs (quasi-skin modes) are still applicable in an open system. Indeed, quasi-skin modes can keep their spatial structure in the open lattice up to a survival period, $\tau$. We examine how system parameters impact this survival time in the following steps.
\\
For the semi-infinite lattice, numerical solutions of the nonlinear equation (\ref{stationary SE}) can be found iteratively. In this method, starting from arbitrary values $ \psi_1 $ and $ \omega $, the solutions, $ \psi_n $, can be given as an $n$-th term of a sequence of relation (\ref{stationary SE}) where we demand that $\psi_{n}\rightarrow 0$ when $n\rightarrow\infty$. Due to the nonlinear nature of the problem, the solutions obtained by the recursion method are potentially unstable since an arbitrarily small change to initial values $ \psi_1 $ and $ \omega $ causes enormous changes in the sequences \cite{Yuce2021PLA}. We are interested in stationary solutions that are stable against perturbation among those found using the iterative method. In other words, for certain values of $ \psi_1 $ and $ \omega $, we are looking for a class of solutions with a particular frequency domain in which the field amplitudes remain bounded and eventually converge to zero. From a technical perspective, they correspond to those solutions of (\ref{stationary SE}) whose absolute values decrease monotonically such that $ \vert\psi_{n+1}\vert<\vert\psi_n\vert $, which guarantees the SIBCs. These solutions are localized at the left edge. Applying this condition leads to
\be
0<\left| \frac{\psi_{n+1}}{\psi_{n}}\right| = \left| \frac{\omega}{\kappa_{R}}-\frac{g}{\kappa_{R}}\vert\psi_n\vert^2-\frac{\kappa_{L}}{\kappa_{R}}\frac{\psi_{n-1}}{\psi_{n}}\right|<1, \label{localized mode relation}
\ee
in light of the reverse triangle inequality, $\vert a-b\vert\geq\vert\vert a\vert-\vert b\vert \vert$, the above inequality leads to
\be
\left| \left|  \frac{\omega+\gamma}{\kappa_{R}}-\frac{g}{\kappa_{R}}\vert\psi_{n}\vert^{2}\right|-\left|\frac{\kappa_{L}}{\kappa_{R}}\frac{\psi_{n-1}}{\psi_{n}}\right|\right|<1
\ee
from this inquality, we get
\be
-1-\frac{\kappa_{L}}{\kappa_{R}}\left| \dfrac{\psi_{n-1}}{\psi_{n}}\right|\leq\frac{\omega}{\kappa_{R}}-\frac{g}{\kappa_{R}}\vert\psi_{n}\vert^2 \leq 1+\frac{\kappa_{L}}{\kappa_{R}}\left| \dfrac{\psi_{n-1}}{\psi_{n}}\right| .
\ee
One can go one step forward to determine the upper and lower bounds of $ \omega $ by considering that for a localized mode, $ \vert\psi_{n-1}/\psi_{n}\vert\geqslant 1 $ as $n\rightarrow\infty$. Then, the relation (\ref{localized mode relation}) is satisfied if
\be
-(2\kappa-g\vert\psi_{1}\vert^2)\leq \omega \leq 2\kappa,\label{upper lower bound}
\ee
where the parameter $\lambda$ is canceled out in both upper and lower limits, which makes them independent of the nonreciprocity parameter. In order to implement skin mode solutions, it is necessary to have a non-zero nonreciprocity parameter.  Nevertheless, the fact that is not present in Eq. 6 suggests that it does not define the possible range for these solutions.\\
Relation \ref{upper lower bound} defines a range of $ \omega $ such that Eq. (\ref{stationary SE}) admits stationary solutions localized at the left edge. Modes with $ \omega>2\kappa $ or $ \omega<g\vert\psi_{1}\vert^2-2\kappa $ are not considered part of the skin mode category. This is because, instead of increasing, these modes remain constant for large values of $ n $ or their amplitude increases significantly. It reveals that for a fixed value, $ \psi_1 $, frequencies of such skin modes are determined by the nonlinear coefficient $ g $ and the coupling strength $ \kappa $.\\
One can find another class of solutions by assuming that $ \vert\psi_n\vert $ increases for a few sites and then decreases monotonically. This happens specifically when the lower bound is positive (e.g., $ g\vert\psi_1\vert^2>2\kappa $), then, for a small $ \lambda $, the lower bound of relation \ref{upper lower bound} fails to provide a condition for having skin mode. To include this class of skin mode, we need to extend the lower bound of (\ref{upper lower bound}) such that
\be
-(2\kappa-g~\text{max}\left\lbrace \vert\psi_n\vert^2\right\rbrace )\leq(\omega+\gamma).\label{lower bound}
\ee
We demonstrate the density of four quasi-skin modes corresponding to various frequencies in Fig. (\ref{SS for Semi-Infinite}). This diagram shows that for $ \omega>2 $ (dashed blue line), which corresponds to those frequencies above the upper bound in Eq. (\ref{upper lower bound}), the skin mode disappears. We observe that for the positive lower bound, there are solutions in which the wave amplitude $ \vert\psi_n\vert $ increases for a few sites before monotonically decreasing. We plot one of these solutions (blue line) in Fig. (\ref{SS for Semi-Infinite}). In this case,  for the lower bound, the relation (\ref{lower bound}) must be considered. These modifications, however, do not give sufficient criteria for small $ \lambda $. We found that the solution blows up at $ \lambda<0.5 $ for a given value of the diagram (\ref{SS for Semi-Infinite}). If we reduce the nonlinearity strength of the system, such that $ g\vert\psi_1\vert^2<2\kappa $, the blow-up solutions disappear. In the next section, we investigate the dynamics of this solution in an open finite lattice.
\section{Stability of quasi-skin mode in a finite lattice}\label{sec2}
\begin{figure*}[t]
	\centerline{\includegraphics[width=1\textwidth]{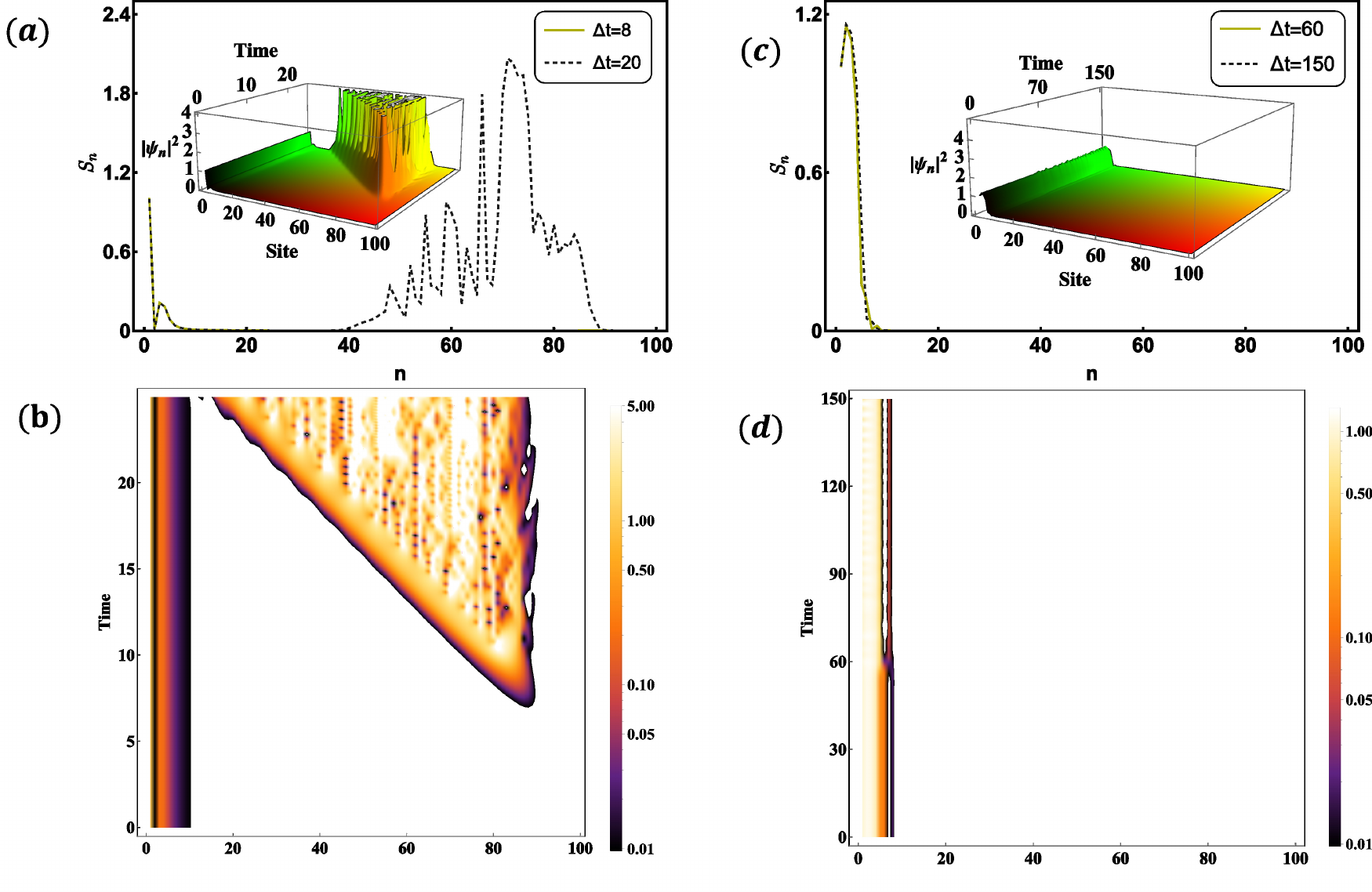}}
	\caption{Diagrams (a,c) display time average $ S_n $ of the amplitude defined by (\ref{time average}) for $ \omega=1.95 $ (left) and $ \omega=0.49 $ (right). The yellow line and dashed black line represent different average time for each frequency. Figs. (b,d) demonstrate the time evolution of amplitude $ \vert\psi_n\vert^2 $. Diagram (a) reveals that the skin mode with the frequency close to the upper bound loses its initial shape after $ \tau=8$. This observation becomes more apparent when examining the time average $S_{n}$ for two distinct time intervals $ \Delta t=8 $ and $ \Delta t=20 $. In the latter, there is a significant mismatch
		between the time average for $ \Delta t=20 $ and the initial profile at the right edge. In Figures. (c,d), we show the time average $S_{n}$ for the localized mode with ($ \omega=0.49 $). As one can see from the diagrams, the initial nonlinear skin mode maintains its profile configuration up to $ \tau=60 $. Based on our numerical findings, the deviation from the initial value for the localized mode with ($ \omega=0.49 $) is negligible even for a longer period. In this study, we scaled parameters such as $t$ representing dimensionless time.}
	\label{P2}
\end{figure*}
We studied the existence of skin modes for a semi-infinite lattice in the previous section. In general, under SIBCs, these stationary solutions do not fulfill OBCs. Nonetheless, up to a survival time, they can be employed as quasi-skin solutions for a finite lattice with OBCs. This section's goal is to compute this survival time and study its relationship to system parameters. Indeed, we numerically find the time evolution of a skin mode in a truncated finite lattice by considering the quasi-skin mode from previous section as an initial condition. To analyze the stability of the nonlinear solutions and to reveal a more quantitative structure of the amplitude, we define the site-dependence time average $S_n$ over a time span from $ T_0 $ to $ T_0+\Delta T $ given by
\be
S_n(\Delta T)=\frac{1}{\Delta T}\int_{T_{0}}^{T_{0}+\Delta T}\vert\psi_n(t)\vert^2dt,\label{time average}
\ee
\begin{figure}
	\begin{center}
		\includegraphics[scale=.35]{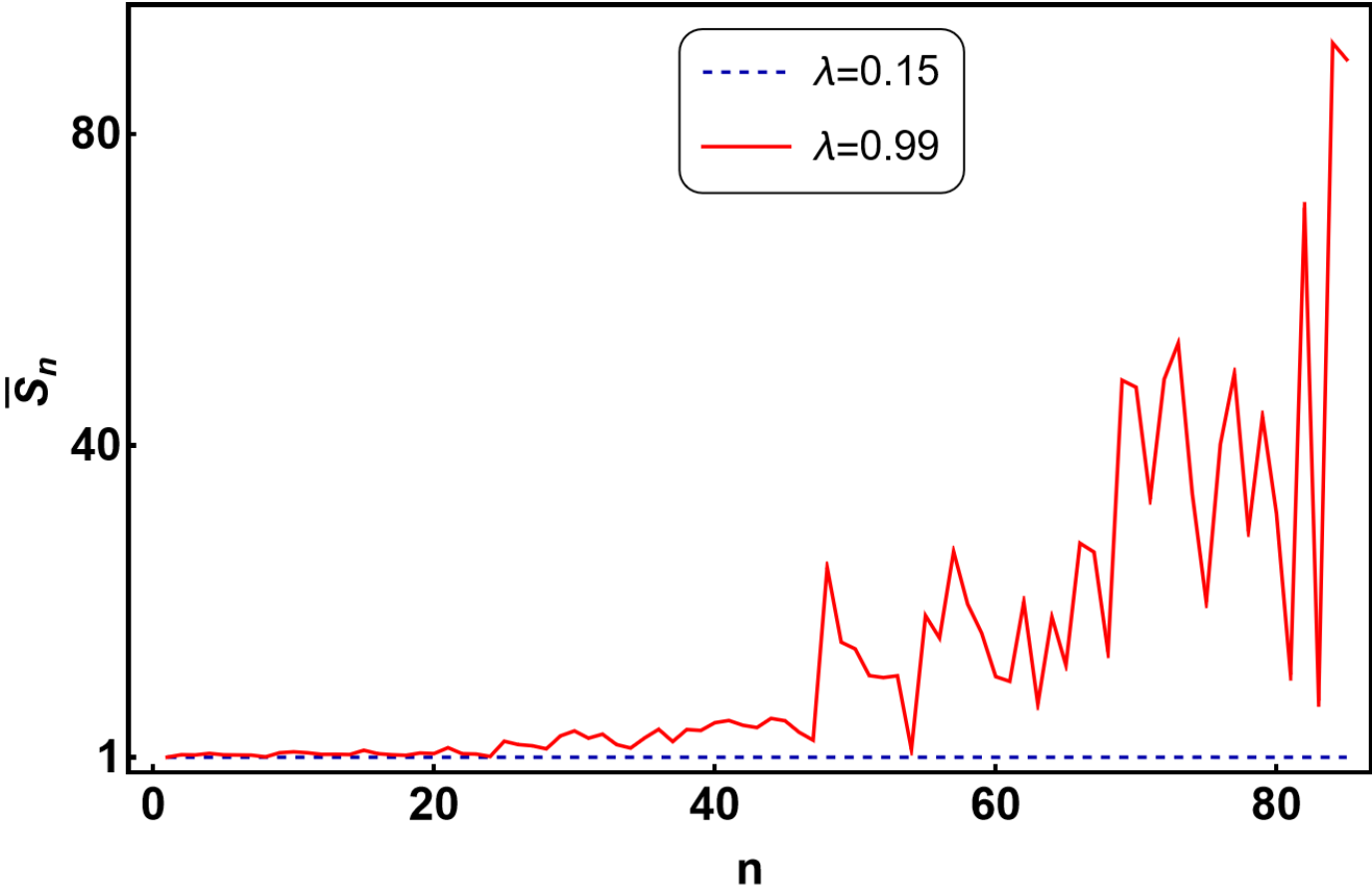}\hspace{0.5cm}
		\caption{The relative site-dependant time average $ \bar{S}_n $ of the dynamical solution of nonlinear Schr\"{o}dinger equation (\ref{dnamical SE}) under open boundary conditions with initial values given by the stationary localized solution of (\ref{stationary SE}) with $ \omega=1.95 $, $ g=2.1 $. The dashed blue line represents the time average for weak nonreciprocity, whereas the solid red line represents the time average for strong nonreciprocity.}
		\label{P3}
	\end{center}
\end{figure}
where $ T_0 $ is our starting point and $ \Delta T $ is the time interval over which we wish to investigate the stability of the wave function. This function provides a measurement for observing the evolution of nonlinear quasi-skin mode in an interval $\Delta T$ \cite{Ezawa2022PRB}. Indeed, for a stable quasi mode, the integrand is almost constant in time (i.e., $\psi_n(t)\approx\psi_n(0)$ for $t\in[T_0,T_0+\Delta T]$), then the function $S_n$ coincides with the quasi mode at initial time $T_0$.\\
Here, we consider two quasi-skin modes with frequencies $\omega=1.95$ and $\omega=0.49$, which are respectively close to the lower and upper bound in relation \ref{upper lower bound}. These two solutions are depicted by the blue and red lines in Fig. \ref{SS for Semi-Infinite}. In Figs. \ref{P2}, we plot the time average $ S_n $ and the time evolution of both solutions. The nonlinear skin mode begins deviating from the initial profile near the right edge for the localized mode with a frequency close to the upper bound. On the other side, the skin mode is more stable for the frequency chosen from the lower bound.\\
In our numerical analysis, we observe that the strength of nonlinearity has no effect on the survival time $ \tau $ of quasi-skin modes. In more detail, the nonlinear parameter defines the frequency range in which an initial quasi skin solution exists. As nonlinearity increases, the quasi-skin mode becomes more localized at the left edge. However, it does not affect this solution's survival time in a finite lattice under open boundary conditions. Interestingly, this is not the case for the nonreciprocity parameter $ \lambda $, which impacts survival time. As previously stated, with the semi-infinite boundary conditions, this parameter does not directly define the possible frequency range for quasi-skin mode. However, the system's dynamics are sensitive to nonreciprocity's strength. To show this, we study the dynamics of a quasi-skin mode with $ \omega=1.95 $ (red line in Figs. (\ref{SS for Semi-Infinite})) in a different range of nonreciprocity. We choose two distinct values for $ \lambda $, one for the weak nonreciprocity ($ \lambda=0.15 $) and the other representing the strong nonreciprocity ($ \lambda=0.99 $). In order to analyze the relation between survival time $\tau$ and the nonreciprocity parameter, we are looking at the relative site-dependant time average given by $\bar{S}_n:=S_n/\vert \psi_n(0)\vert$.
This parameter measures stability by ensuring that the relative time average $\bar{S}_n$ is one for a stable quasi-skin mode within the time interval $ \Delta T$. Alternatively, one may define the survival time by considering both an upper and lower bound of $\bar{S}_{n}$. For example, the survival time is an interval in which the value of $\bar{S}_{n}$ falls within the range of $[1-\epsilon,1+\epsilon]$, where $\epsilon$ is a small parameter. In Fig. \ref{P3}, we plot $ \bar{S}_n $ in the time interval $ \Delta T= 25 $ for two different values of nonreciprocity. The graphs show that in the case of weak nonreciprocity $\lambda=0.15$, the relative time average remains constant for all lattice sites throughout the time interval $\Delta T=25 $, proving the stability of the nonlinear quasi-skin mode during this period. On the other side, in the almost unidirectional lattice ($ \lambda=0.99 $), the profile of the initial quasi-skin mode is distorted near the right edge in the same time interval. We performed the same numeric analysis for the skin solution with $ \omega=0.49 $ and found the same effect of nonreciprocity parameter $ \lambda $ on the survival time. This indicates that large nonreciprocity amplifies the effect of nonlinearity in making the quasi-skin mode unstable, while the initial quasi-skin mode is more localized in a unidirectional system or a system with strong nonreciprocity, and it lives longer in a lattice with weak nonreciprocity.\\
Skin mode's life is also affected by the size of the open system, as it lives longer in a larger lattice. Indeed, as the number of sites grows, the system approaches the semi-infinite boundary conditions, and the original wave packet approaches the stationary solution.\\
\begin{figure*}[t]
	\centerline{\includegraphics[width=1\textwidth]{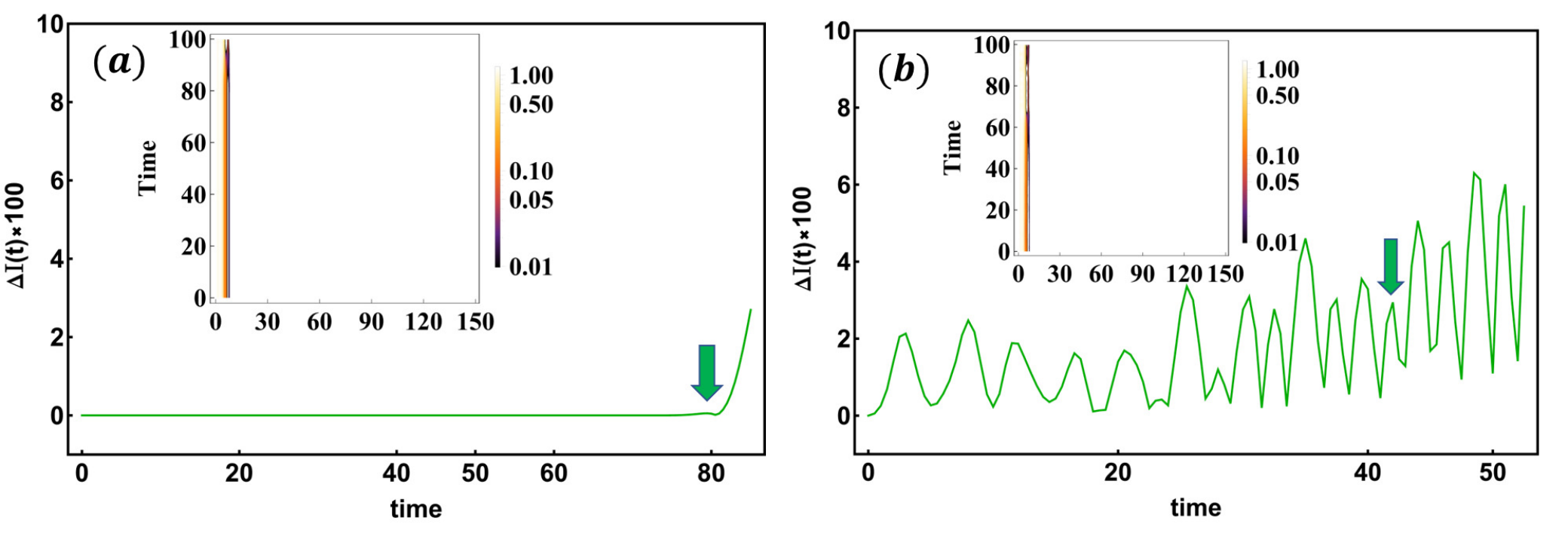}}
	\caption{Diagrams depict the intensity deviation for skin mode with a frequency $\omega=0.49$ in both a noiseless system (on the left) and in the presence of the weak noise (on the right). Diagram (b) illustrates that in the presence of noise, the deviation is less than four percent (green arrow), up to half of the survival time in the absence of noise. In diagram (b), we set the noise strength ($w=10^{-2}$), which can represent an unavoidable noise throughout the system.}
	\label{P4}
\end{figure*}
To study the effect of disorder on $\tau$, we consider defective left and right couplings respectively given by $\kappa_{L}+\delta\kappa_{L}$ and $\kappa_{R}+\delta\kappa_{R}$, where $\delta\kappa_{L, R}$ chosen randomly from the interval $w[-0.5,0.5]$ with the disorder strength $w$. One can intuitively expect that increasing the disorder's strength will decrease the survival time. To show this, we numerically evaluate the intensity deviation given by $\Delta I(t)=\vert(I(t)-I(0))/I(0)\vert$ where $I(t):=\sum_{n=1}^{N}\vert\psi_{n}(t)\vert^2,$
is the total power. Our finding reveals that for an unavoidable defect, one can find a skin mode that keeps its form up to a considerable survival time in the presence of noise. In Fig. \ref{P4}, for a quasi-skin mode with frequency $\omega=0.49$, we plot $\Delta I(t)$ for the noiseless system (i.e., $w=0$) and the weak noise (i.e., $w\approx 10^{-2}$). As one can see from the diagrams, by adding noise, the fluctuations are less than four percent up to around $t=20$, which is half of the survival time $\tau$ of the noiseless system but still considerable. We can boost the stability in the presence of noise by extending the lattice.
\section{Conclusions}
The linear non-Hermitian skin effect nowadays is a hot topic in topological systems whose many aspects have been explored in theoretical and experimental views point. However, the analytic investigation of its nonlinear counterpart is almost novel. This paper studies the existence and survival of quasi-skin mode in a truncated finite lattice. In the first step, we explored the upper and lower bound of frequency in which one can have a class of stationary skin solution for a nonlinear nonreciprocal lattice under semi-infinite boundary conditions. Our approach reveals that the lower bound depends on the nonlinear parameter such that there is no skin mode in low frequency for strong nonlinearity. The nonlinear solutions of a semi-infinite lattice do not satisfy OBCs in general. Nevertheless, they can still survive approximately in a lattice with open edges up to a survival time. Here we found this survival time and explored the effect of the system's parameters on it. While nonlinearity's strength does not affect skin mode's survival time, nonreciprocity plays an important role. We observed that for strong nonreciprocal couplings, the localized mode does not survive for a long time. In the end, we analyzed the survival time in the presence of the noise. Our numeric computations reveal that for an unavoidable noise, one can still obtain quasi-skin mode with a considerable survival time. Our research offers valuable understanding of a new method for creating a consistent skin mode in a nonlinear non-hermitian lattice and analyzing its behavior under physical conditions. This has potential implications in nonlinear topological photonics systems \cite{nonlineartopological}. This work examines the nonlinear version of nonreciprocal tight-binding models, focusing on the analysis of nonlinear skin modes. Further investigations can be conducted in future study to explore other non-linear topological phenomena, such as non-Hermitian edge bursts.

\appendix

\end{document}